\documentclass[superscriptaddress,showpacs,preprintnumbers,amsmath,amssymb]{revtex4}

\usepackage{epsfig}
\usepackage{dcolumn}
\usepackage{bm}
\usepackage{latexsym}

\usepackage{dcolumn}
\usepackage{amsmath}
\usepackage{amssymb}
\usepackage{bm}
\usepackage{color}

\def\be{\begin{equation}}
\def\ee{\end{equation}}
\def\bea{\begin{eqnarray}}
\def\eea{\end{eqnarray}}

\begin{document}
\preprint{APS/123-QED}

\title{Magnetic field control of photon echo in the electron-trion system:\\ Shuffling of coherences between optically accessible and inaccessible states}

\author{L. Langer}
 \affiliation{Experimentelle Physik 2, Technische Universit\"at Dortmund, 44221 Dortmund, Germany}
\author{S.~V. Poltavtsev}
 \affiliation{Experimentelle Physik 2, Technische Universit\"at Dortmund, 44221 Dortmund, Germany}
 \affiliation{Spin Optics Laboratory, St. Petersburg State University, 198504 St. Petersburg, Russia}
\author{I.~A. Yugova}
 \affiliation{Experimentelle Physik 2, Technische Universit\"at Dortmund, 44221 Dortmund, Germany}
 \affiliation{Spin Optics Laboratory, St. Petersburg State University, 198504 St. Petersburg, Russia}
\author{D.~R. Yakovlev}
 \affiliation{Experimentelle Physik 2, Technische Universit\"at Dortmund, 44221 Dortmund, Germany}
 \affiliation{A.F. Ioffe Physical-Technical Institute, Russian Academy of Sciences, 194021 St. Petersburg, Russia}
\author{G.~Karczewski}
\author{T. Wojtowicz}
\author{J. Kossut}
 \affiliation{Institute of Physics, Polish Academy of Sciences, PL-02668 Warsaw, Poland}
\author{I.~A. Akimov}
 \affiliation{Experimentelle Physik 2, Technische Universit\"at Dortmund, 44221 Dortmund, Germany}
 \affiliation{A.F. Ioffe Physical-Technical Institute, Russian Academy of Sciences, 194021 St. Petersburg, Russia}
\author{M. Bayer}
 \affiliation{Experimentelle Physik 2, Technische Universit\"at Dortmund, 44221 Dortmund, Germany}
\date{\today}

\begin{abstract}
We report on magnetic field induced oscillations of the photon echo signal from negatively charged excitons in a CdTe/(Cd,Mg)Te semiconductor quantum well. The oscillatory signal is due to Larmor precession of the electron spin about a transverse magnetic field and depends sensitively on the polarization configuration of the exciting and refocusing pulses. The echo amplitude can be fully tuned from maximum down to zero depending on the time delay between the two pulses and the magnetic field strength. The results are explained in terms of the optical Bloch equations accounting for the spin level structure of electron and trion.
\end{abstract}




\maketitle

Coherent optical phenomena in ensembles of atoms or other systems with discrete energy levels have attracted considerable attention in relation to optical quantum memories \cite{Hammerer10, Lvovski09}. One important phenomenon in this respect is the photon echo where in a classical picture an intense optical pulse results in rephasing and retrieval of a macroscopic medium's polarization that was created by a preceding optical pulse. The extension by a third pulse may even stimulate the photon echo such that the retrieval occurs on demand \cite{Mukamel}. An inherent feature for emergence of a photon echo is presence of an inhomogeneity in the system, by which the optically imprinted phase information is spread in the ensemble while still being maintained by the individual constituents. Such inhomogeneity is an unavoidable property for many systems of interest. Current experimental activities on quantum light-matter interfaces focus on alkali atoms or impurity centers in rare earth ions or diamond defects \cite{Hammerer10}. Different protocols using controlled reverse inhomogeneity or atomic frequency combs have been developed in order to overcome noise in conventional photon echo schemes and to increase the optical memory efficiency for single photon applications \cite{Lvovski09, Afzelius10,Tittel11}.

So far, semiconductor nanostructures were not considered as prime candidates mainly due to high decoherence rates and complex band structures. However, compared to isolated atoms the fundamental optical excitation in semiconductors, the exciton, possesses a large dipole moment allowing fast operation. Moreover its large absorption strength leads to high efficiency, which is a prerequisite for operation with single photons. On the other hand there is some tradeoff as a large dipole moment also shortens the exciton lifetime (down to about a nanosecond as compared to possible milliseconds in atomic systems). This is an important obstacle since the memory should be stored long enough that, e.g., after reconversion into a photon this photon might be interfered with another photon arriving at a later time. In this respect charged excitons (trions) open new possibilities in which the associated long-lived electron or hole spin in the ground state can be exploited \cite{SeminoductorQubits,SpinPhysics}. In quantum dots (QDs) where the spin-orbit interaction is strongly suppressed the electron spin lifetime may be as long as milliseconds \cite{Finley04}. Significant progress in coherent optical control of electron spin in charged QDs has been demonstrated by several groups \cite{Yamamoto08,Awschalom08,Xu08,Greilich09}. Moreover, spin frequency combs can be prepared using, for example, electron spin mode-locking \cite{mode-locking}. Thus an ensemble of trions is prospective for applications in optical memories.

Our concept relies on the photon echo in the electron-trion system subject to transverse magnetic field. The transverse field plays a crucial role here, as it involves the electron spin into the echo because of Larmor precession: the magnetic field introduces a coupling between the spin states along the optical axis that cannot be induced by the optical field. This coupling allows transfer of a coherent superposition between a pair of states that is optically accessible (hereforth termed "bright" coherence) into a superposition between a pair of states that is optically inaccessible ("dark" coherence). In particular transfer of the bright coherence into a long lived, dark electron spin coherence is possible. In what follows we first describe the main idea of this magnetic field control and then present the experimental demonstration for trions in a semiconductor CdTe/(Cd,Mg)Te quantum well (QW).

\begin{figure*}
\epsfxsize= 17.5 cm
\centerline{\epsffile{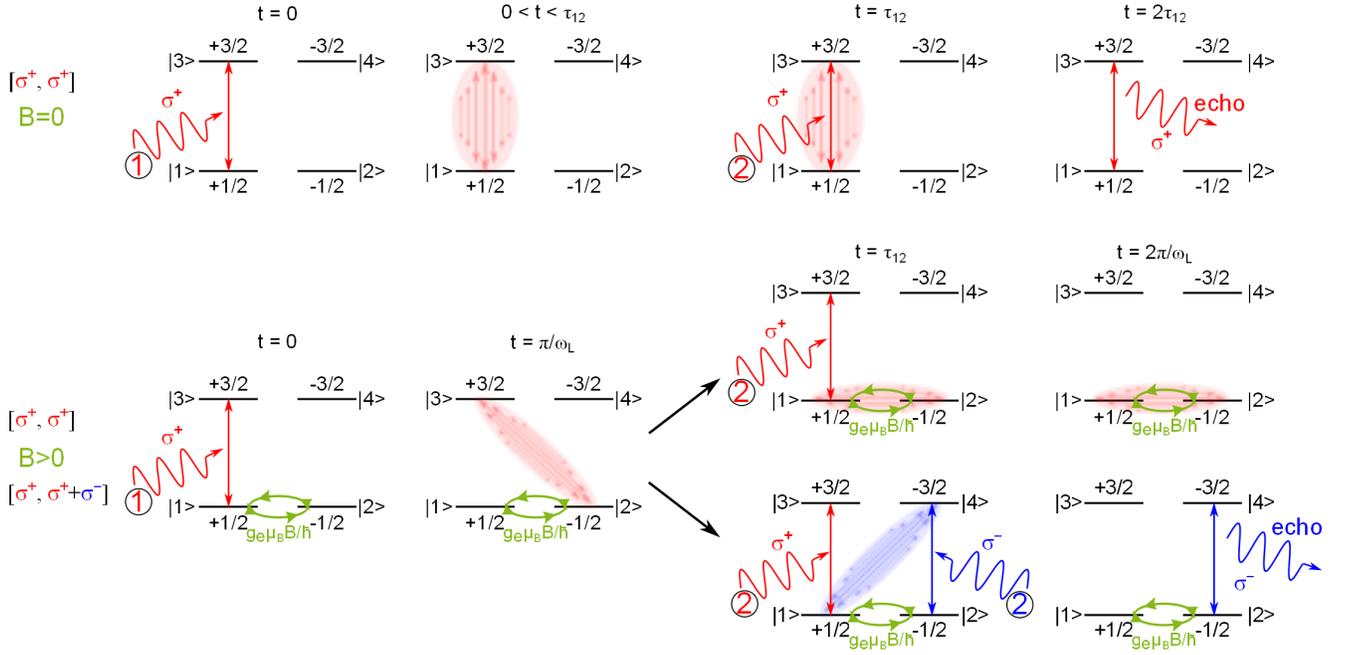}}
\caption{\label{fig:schema} (Color online) Schematic representation of photon echo in the electron-trion system. Time $t$ evolves from left to right. Three different scenarios are considered. The top line shows the photon echo in the two-level system of the $|1\rangle$ and $|3\rangle$ states ($B=0$,$[\sigma^+,\sigma^+]$). The mid line shows the transfer of optical coherence into electron coherence ($B>0$,$[\sigma^+,\sigma^+]$), so that no photon echo occurs. The bottom line shows the transfer of optical coherence from the left $|1\rangle - |3\rangle$ two-level system accessible by $\sigma^+$-polarized light into the $|2\rangle - |4\rangle$ system addressable by $\sigma^-$-polarized light. The photon echo is correspondingly $\sigma^-$ polarized.}
\end{figure*}

The energy and spin level structure of the electron-trion system are schematically shown in Fig.~\ref{fig:schema}. The trion complex comprises two electrons and a hole. In the trion ground state the electron spins are antiparallel to each other and therefore the trion is a doublet state according to the total angular momenta of the hole $J=3/2$. This situation holds not only for a QW, but also for a quantum dot. Optical excitation couples the electron with possible spin projections $S_z=\pm1/2$ and the trion doublet with $J_z=\pm3/2$, respectively, where the quantization $z$-axis is perpendicular to the QW or QD plane. Both doublets are half integer spin states and hence their levels are degenerate in absence of an external magnetic field. The electric-dipole selection rules allow two optical transitions $|1\rangle + \sigma^+ \rightarrow  |3\rangle$ and $|2\rangle + \sigma^-  \rightarrow |4\rangle$, where $\sigma^+$ and $\sigma^-$ denote the respective circular photon polarization. The system can be considered as a pair of uncoupled oscillators [the left and right columns of the four-level-scheme in Fig.~\ref{fig:schema} are independent]. Application of a transverse magnetic field leads to coupling between the electron spin states $|1\rangle$ and $|2\rangle$ and the trion states $|3\rangle$ and $|4\rangle$. For the trions this effect is small because the in-plane $g$-factor of the heavy-hole is about zero. Therefore the main contribution comes from the electron spins with Larmor precession frequency $\omega_L=g_e\mu_B B/\hbar$ around the magnetic field ($\mathbf{B} \| x$). Here $g_e$ is the electron g-factor and $\mu_B$ is the Bohr magneton.

We consider a sequence of two optical pulses tuned in resonance with the electron-trion transition and propagating along the $z$-axis (Voigt geometry). Pulse 2, the refocusing pulse, is delayed by time $\tau_{12}$ in respect to pulse 1 (the excitation pulse). We operate with short pulse approximation where the pulse duration $t_p$ is significantly shorter than the trion lifetime, the decoherence time and the electron spin precession period ($\omega_Lt_p\ll1$). Therefore we can separate the interaction of the electron-trion system with light from its dynamics in magnetic field.

Characteristic scenarios for excitation with a sequence of exciting and refocusing pulses are shown in Fig.~\ref{fig:schema} for different polarization configurations. For simplicity we neglect decoherence processes here. The top line is at $B=0$ for $\sigma^+$ polarized pulses ([$\sigma^+,\sigma^+$] configuration). At $t=0$ optical coherence between the states $|1\rangle$ and $|3\rangle$ is created by the first pulse (solid red arrow). Due to inhomogeneity of the optical transitions dephasing takes place so that the macroscopic coherence disappears. This is indicated by the set of arrows of different lengths symbolizing the phase distribution for the dipoles excited with different frequencies. At $t=\tau_{12}$ pulse 2 conjugates the phase for each dipole and rephasing starts. At time $t=2\tau_{12}$ the rephasing has been completed. This results in a photon echo, which is emitted with the same polarization as the two pulses. Note that the photon echo can occur only if both pulses are co-polarized. This scenario would also take place in a two level system with states $|2\rangle$ and $|4\rangle$ for a sequence of $\sigma^-$ polarized pulses.

At $B>0$ the ground state electron precesses (green arrows in Fig.~\ref{fig:schema}). The different coherences showing up now between states $|i\rangle$ and $|j\rangle$ can be associated with non-diagonal elements of the density matrix $\rho_{ij}, i\neq j$. After half of a Larmor precession period, $t=\pi/\omega_L$ the \textit{optical} coherence $\rho_{13}$ has been fully shuffled to $\rho_{23}$. The latter represents a "dark" coherence. No photon echo can occur if the system is exposed to a circularly polarized pulse at this point of time. Nevertheless the second pulse induces the transition from $|1\rangle$ to $|3\rangle$, which leads to transfer of coherence from $\rho_{23}$ to $\rho_{21}$ [see  the second line in Fig.~\ref{fig:schema}]. The latter matrix element corresponds to a long-lived electron spin. The microscopic coherence is frozen and stored in the electron spin, i.e. no dephasing or rephasing will occur further, only Larmor precession.

Another interesting scenario takes place when the refocusing pulse is linearly polarized $H=(\sigma^++\sigma^-)/\sqrt{2}$, i.e. contains  $\sigma^+$ and $\sigma^-$ polarizations (see third line of Fig.~\ref{fig:schema}). In this case $\rho_{23}$ is transferred to $\rho_{14}$, which is also a "dark" coherence but due to Larmor precession it is finally shuffled into $\rho_{24}$ at $t=2\tau_{12}$. In addition rephasing has been also accomplished at this point of time so that a $\sigma^-$ polarized photon echo is observed.

The scenarios described above are only examples for understanding the main principles of the magnetic field effect on the photon echo. The complete picture can be developed by solving the Lindblad equation of motion for the ($4\times4$) density matrix $\rho_{ij}(t)$ of the four level electron-trion system of Fig.~\ref{fig:schema} (see Appendix A). We assume that initially at $t=0$ the electron spins are unpolarized, $\rho_{11}(0)=\rho_{22}(0)=1/2$, while all other elements of the density matrix are zero. In addition we consider optical pulses of rectangular temporal profile. The main contributions are summarized in Table \ref{tab:config}. All other polarization configurations result from linear combination of these contributions. For example, if we consider a sequence of two linearly co-polarized pulses there is no influence of the magnetic field on the photon echo amplitude. However, if we apply two linearly cross-polarized pulses the photon echo shows oscillatory behavior. For example in the [$V$,$H$] configuration the photon echo is $V=(\sigma^+-\sigma^-)/\sqrt{2}$ polarized with an amplitude given by $P_{pe}^V \propto \cos{(\omega_L\tau_{12})}$. Obviously $P_{pe}^V$ then changes sign at $\omega_L\tau_{12}=\pi$.

\begin{table}[b]
\caption{\label{tab:config}%
Non-zero contributions to the photon echo signal for different excitation polarization configurations.}
\begin{ruledtabular}
\begin{tabular}{ccc}
&\multicolumn{2}{c}{photon echo polarization}\\
polarization \\ configuration & $\sigma^+$ & $\sigma^-$ \\
\hline\\
$\sigma^+\sigma^+$ & $\cos^2(\omega_L\tau_{12}/2)$ & 0 \\
$\sigma^-\sigma^-$ & 0 & $\cos^2(\omega_L\tau_{12}/2)$  \\
$\sigma^+H$ & $\frac{1}{2}\cos^2(\omega_L\tau_{12}/2)$ & $\frac{1}{2}\sin^2(\omega_L\tau_{12}/2)$ \\
$\sigma^-H$ & $\frac{1}{2}\sin^2(\omega_L\tau_{12}/2)$ & $\frac{1}{2}\cos^2(\omega_L\tau_{12}/2)$ \\
\end{tabular}
\end{ruledtabular}
\end{table}

\begin{figure*}
\epsfxsize= 17.5 cm
\centerline{\epsffile{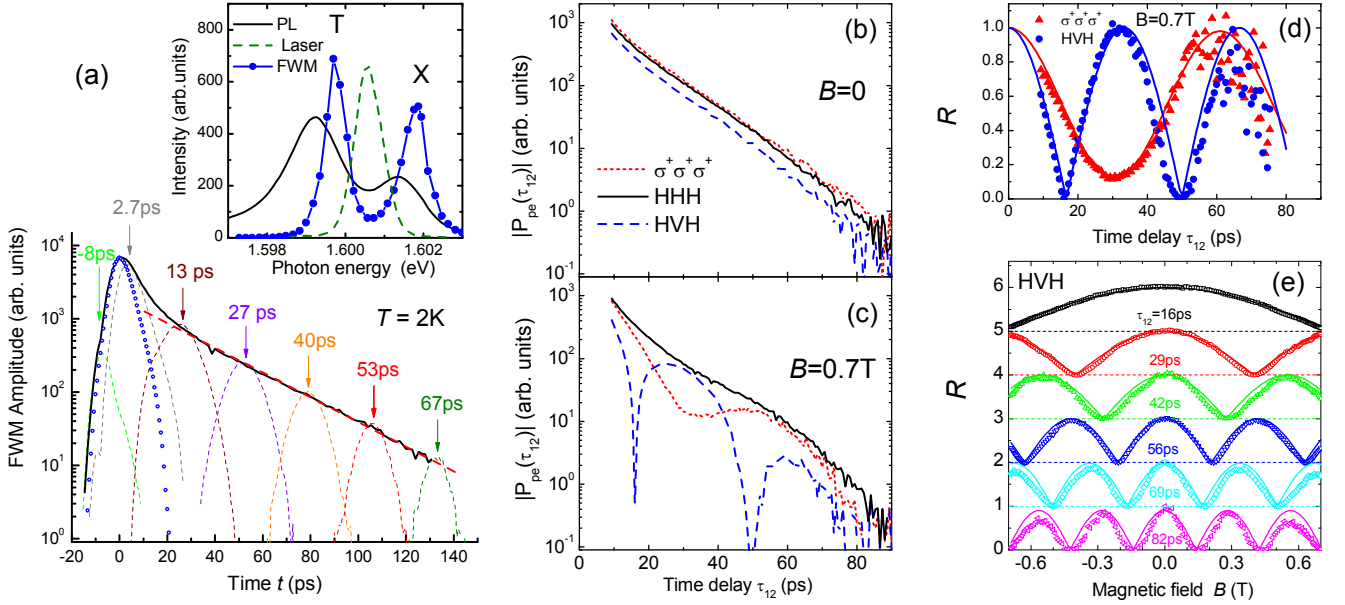}}
\caption{\label{fig:experiment} (Color online) (a) Demonstration of photon echo. Dashed curves correspond to time-resolved FWM amplitudes measured for different time delays $\tau_{12}$, as indicated also at the arrows. Open symbols as well as solid line give the FWM signal decay at $t=\tau_{12}$ and $t=2\tau_{12}$, respectively. $t=0$ corresponds to pulse 1 arrival time. Dashed line is fit by exponential decay. Inset: spectra of FWM signal measured at $t=\tau_{12}=0$, PL and laser. PL spectrum is measured for cw excitation with photon energy 2.33~eV. (b) and (c) Photon echo amplitude as function of $\tau_{12}$ for three different excitation-detection polarization configurations $HHH$, $HVH$ and $\sigma^+\sigma^+\sigma^+$ at $B=0$ (b) and $B=0.7$~T (c). (d) Ratio $R$ as defined in text as function of $\tau_{12}$ in the $HVH$ (circles) and $\sigma^+\sigma^+\sigma^+$ (triangles) configurations. (e) Magnetic field dependence of $R$ in $HVH$ configuration. Solid lines in (d) and (e) are theory curves according to Table~\ref{tab:config}. The oscillation frequency $\Omega$ is determined by $|g_e|$=1.67 in $\sigma^+\sigma^+\sigma^+$ and $|g_e-g_{h,\bot}|$=1.54 in $HVH$ configuration. }
\end{figure*}

To address the trion ensemble only, we perform the experiments on a semiconductor QW structure for which the neutral and charged (trion) excitons are spectrally well separated. The energy level structure of a QW trion is very similar to that in a QD when the trion is localized as typically is the case at cryogenic temperatures. Only its coherence time is shorter in the QW. The investigated sample comprises 5 decoupled 20~nm thick CdTe QWs separated by 110~nm Cd$_{0.78}$Mg$_{0.22}$Te barriers. Information about electronic and optical properties, e.g. the electron and hole g-factors, spin dephasing and radiative decay times, can be found in \cite{Zhukov06, Bartsch11}.

Transient four-wave mixing (FWM) allows direct investigation of coherence in semiconductors ~\cite{Chelma01}. So far only a few transient FWM experiments have been done on semiconductor structures in magnetic field \cite{Bar-Joseph93, Chelma95, Cundiff96, Kner98, Davis07}. Many body interactions between the distinct magnetoexciton states and their Fano interferences were demonstrated using spectrally broad laser pulses in magnetic fields up to 10~T \cite{Chelma95, Cundiff96, Kner98}. Here we use spectrally narrow picosecond pulses to excite only the trion states and minimize many-body interactions. In addition the experiments are performed at low power densities (pulse energy below 100~nJ/cm$^{2}$), corresponding to the linear excitation regime for each pulse. Interferometric heterodyne detection provides high sensitivity for measuring the absolute value of the FWM electric field amplitude in real time. All measurements were performed at temperature $T=2$~K. Details about the experimental technique are described in Appendix B.

The spectral dependence of the FWM signal measured at zero time delay $t=\tau_{12}=0$ is shown in the inset of Fig.~\ref{fig:experiment}(a).  The PL spectrum coincides well with the FWM signal except of a Stokes shift of about 0.5~meV. This shift indicates spectral diffusion of the exciton and trion complexes to localization sites due to fluctuations of QW width and composition. The trion peak T in the FWM spectrum is located at 1.5997~eV, and is red shifted relative to the exciton X by the binding energy of 2.2~meV. The time-resolved FWM amplitude measured at the T resonance is shown in Fig.~\ref{fig:experiment}(a). At time delays $\tau_{12} < 10$~ps the form of the signal contains signatures of free polarization decay. At later delays the coherent response is given by a Gaussian like pulse, appearing at a delay time of $t=2\tau_{12}$, thereby representing a photon echo. The amplitude of the photon echo decays exponentially with increasing $\tau_{12}$. The decay is well described by $|P_{pe}| \propto \exp{(-2\tau_{12}/T_2)}$ with a coherence time $T_2=25$~ps, which agrees well with previous studies on similar p-doped QWs \cite{Brinkmann99}.

The decays $|P_{pe}(\tau_{12})|$ for different polarization configurations are summarized in Fig.~\ref{fig:experiment}(b). The configurations are denoted by $ABC$, where $A$ and $B$ are the polarizations of the excitation and refocusing pulses 1 and 2, respectively, and $C$ is the polarization of the resulting photon echo amplitude. For zero magnetic field the temporal behavior of the signal is shown for the $HHH$, $HVH$ and $\sigma^+\sigma^+\sigma^+$ configurations. Here both the coherence times and the amplitudes are nearly the same, indicating that many-body interactions can be neglected \cite{Chelma01}. At $B=0.7$~T strong oscillations occur for the $HVH$ and $\sigma^+\sigma^+\sigma^+$ configurations [see Fig.~\ref{fig:experiment}(c)]. These oscillations appear also when the magnetic field $B$ is scanned.

For convenience we consider the ratio $R=|P_{pe}(\tau_{12},B)|/|P_{pe}(\tau_{12},B=0)|$ to isolate the oscillatory part, assuming that $T_2$ is not altered for magnetic fields up to $B=1$~T. The temporal and magnetic field dependencies of  $R(\tau_{12},B)$ are presented in Figs.~\ref{fig:experiment}(d) and (e). In full agreement with our expectations (see Table~\ref{tab:config}) $R$ follows a $\cos^2(\Omega\tau_{12}/2)$ dependence in the $\sigma^+\sigma^+\sigma^+$ configuration, while for the $HVH$ configuration $R$  oscillates as $|\cos(\Omega\tau_{12})|$ \footnote{For the $\sigma^+\sigma^+\sigma^+$ configuration we take into account a constant background $R_0=0.12$ resulting from residual linear polarization of the laser pulses. A rotation of the photon echo polarization is excluded: Within the experimental error we observe no signal in the $HVV$ configuration.}. For co-circularly polarized pulses the oscillation frequency $\Omega$ corresponds to the electron Larmor precession frequency $\omega_L$ given by $|g_e|=1.67$, while in the $HVH$ configuration we evaluate a somewhat smaller frequency corresponding to $|g_e-g_{h,\bot}|=1.54$. Here Larmor precession of the heavy hole in the trion state contributes to the photon echo oscillation frequency resulting in $\Omega =|g_e-g_{h,\bot}|\mu_B B/\hbar$, where $|g_{h,\bot}|=0.13$ is the transverse heavy-hole g-factor (see Appendix A).

In conclusion we have demonstrated magnetic control of photon echo in the electron-trion system of a semiconductor quantum well. Exploiting the Larmor precession of electron spins in transverse magnetic field we demonstrate transfer of coherence between optically accessible and inaccessible state superpositions. This allows us to suppress the photon echo amplitude or change its sign depending on the magnetic field strength and delay between the exciting and refocusing pulses. The observed signals can be well described by the optical Bloch equations taking into account the spin level structure of electron and trion. Note that the external magnetic field can be substituted by effective magnetic fields produced by the nuclei or magnetic impurities due to hyperfine or exchange interactions.

This work was supported by the Deutsche Forschungsgemeinschaft (Grants No. AK-40/3-1 and AK-40/4-1). I.A.Y. is a Fellow of the Alexander von Humboldt Foundation. The authors are grateful to V.L.~Korenev and I.Ya.~Gerlovin for useful discussions.

\appendix

\section{Photon echo amplitude in different polarization configurations}

To calculate the FWM signal it is necessary to determine the trion
polarization under the action of the two laser pulses, the excitation pulse and the refocusing pulse.
By definition, the medium polarization is the dipole moment of the system,
which in quantum-mechanical representation is given by the mean
value of the dipole moment operator $\hat{d}$:
\begin{equation}
\label{eq:eq1}
P=Tr(\hat{d}\rho)
\end{equation}
here $\rho$ is density matrix of the electron-trion system and $Tr$
is trace of the matrix. The temporal evolution of the density matrix
is described by the Lindblad equation:
\begin{equation}
\label{eq:eq2}
i\hbar \dot{\rho}=[\hat{H},\rho]+\Gamma.
\end{equation}
Here $\hat{H}$ is the Hamiltonian of the system and $\Gamma$
describes relaxation processes phenomenologically. In our case the
Hamiltonian contains three contributions:
$\hat{H}=\hat{H}_0+\hat{H}_B+\hat{V}$, where $\hat{H}_0$ is the
Hamiltonian of the unperturbed spin system, $\hat{H}_B$ gives the
interaction with magnetic field and $\hat{V}$ describes the
interaction with light. In our calculation we use the short pulse
approximation assuming that the pulse duration is significantly
shorter than the trion lifetime, the decoherence times and the
electron spin precession time about a transverse magnetic field.
This assumption is justified for our experimental conditions.

Under these conditions, we can separate and consider consistently
the interaction of the electron-trion system with light and its
dynamics in magnetic field. More specifically, to calculate the FWM
signal we take the following steps: (1) Calculation of the density
matrix during the first pulse action. Here we can neglect relaxation
processes and spin precession. This gives us the density matrix at
the end of the first pulse $\rho (t_p)$ ($t_p$ is the pulse
duration). (2) Calculation of the density matrix, taking into
account precession and relaxation with $\rho (t_p)$ as initial
condition. In this way we get the density matrix at the time of
arrival of the second pulse, $\rho (\tau_{12})$, where $\tau_{12}$
is the delay between pulses. (3) Then, again, one needs to calculate
$\rho$ under the second pulse action, similar to (1), $\rho
(t_p+\tau_{12})$. (4) At last, one calculates the time evolution of
$\rho (t)$ in magnetic field  after the second pulse. Using this
matrix, one can obtain the FWM signal. In the following the calculational procedure of the
FWM signal is detailed and expressions are given for the cases of excitation by circular or
linear polarized pulses with and without taking into account the spin
precession of the hole.

First we consider the effect of photoexcitation by a short laser
pulse with frequency $\omega$ close to the trion resonant frequency
$\omega_0$. We consider a situation with low concentration of
resident carriers in the quantum well and consider resonant (or
almost resonant) optical excitation of trions that do not interact
with each other. We completely neglect all other excited states of
the QW. Apart from that we consider pulses with rectangular shape
for simplicity, which allow us to get analytical solutions for the
density matrix. In addition, this approximation is good enough for
describing of our experimental data because the relevant excitation
regime is the low power regime in terms of pulse area
\cite{rsa_vs_ml}.

The incident electromagnetic field induces optical transitions
between the electron state and the trion state creating a coherent
superposition of these states. In accordance with the selection
rules, $\sigma^+$ circularly polarized light creates a superposition
of the $+1/2$ electron and $+3/2$ trion states, while $\sigma^-$
polarized light creates a superposition of the $-1/2$ electron and
$-3/2$ trion states. In order to describe these superpositions it is
convenient to introduce a 4x4 time dependent density matrix,
comprising the two electron spin projections ($\pm 1/2$) (index 1
and 2) and two hole spin projections ($\pm 3/2$) (index 3 and 4).

The interaction with the electromagnetic wave in the dipole approximation is described
by the Hamiltonian:
\begin{equation}
 \label{eq:eq3}
\hat V(t) = -\int [\hat d_+(\bm r) E_{\sigma^+}(\bm r,t) +
\hat d_-(\bm r)E_{\sigma^-}(\bm r,t)] \mathrm d^3 r\:,
\end{equation}
where  $\hat d_\pm(\bm r)$ are the circularly polarized components
of the dipole moment density operator, and $E_{\sigma ^\pm}(\bm
r,t)$ are the correspondingly polarized components of the electric
field of a quasi-monochromatic electromagnetic wave. The electric
field of this wave is given by
\begin{equation}
\bm E(\bm r, t) = E_{\sigma^+}(\bm r,t) \bm o_+  + E_{\sigma^-}(\bm
r,t)\bm o_- + {\rm c.c.}\:, \label{eq:eq4}
\end{equation}
where $\bm o_\pm$ are the circularly polarized unit vectors that are
related to the unit vectors ${\bm o}_x \parallel x$ and ${\bm o}_y
\parallel y$ through $\bm o_\pm = (\bm o_x \pm \mathrm i \bm
o_y)/\sqrt{2}$. Here the components $E_{\sigma ^+}$ and
$E_{\sigma^-}$  contain temporal phase factors $ \mathrm e^{-\mathrm
i \omega t}$.

The strength of the light interaction with the electron-trion system
is characterized  by the corresponding transition matrix element of
the operators $\hat{d}_{\pm}({\bm r})$ calculated with the wave
functions of the valence band, $|\pm 3/2\rangle$, and conduction
band, $|\pm 1/2\rangle$: \cite{ivchenko05a}
\begin{equation}
\label{dpm}
\mathsf d(\bm r) = \langle 1/2 |\hat d_- (\bm r)|3/2\rangle =
\langle - 1/2 |\hat d_+ (\bm r)|-3/2 \rangle.
\end{equation}

We assume, that the trion recombination time is considerably shorter
than the laser repetition period. Therefore, before the first pulse
only elements of the density matrix belonging to the electron are
unequal zero. We also assume, that these are only populations
$\rho_{11}$ and $\rho_{22}$. The solution of the von Neumann
equation $i\hbar \dot{\rho}=[\hat{H}_0+\hat{V},\rho]$ gives after
the first pulse action: \bea
\rho_{13}(t_p)&=&{f^*_{1+}\rho_{11}(0)\over 2\Omega_{1+}^2}\left[v(1-\cos\Omega_{1+} t_p)+i\Omega_{1+} \sin\Omega_{1+} t_p\right]\mathrm e^{i\omega t_p},\nonumber\\
\rho_{24}(t_p)&=&{f^*_{1-}\rho_{22}(0)\over 2\Omega_{1-}^2}\left[v(1-\cos\Omega_{1-} t_p)+i\Omega_{1-} \sin\Omega_{1-} t_p\right]\mathrm e^{i\omega t_p},\nonumber\\
\label{eq:eq12} \eea where $t_p$ is the pulse duration.
$\rho_{11}(0)$, $\rho_{22}(0)$ are the initial state populations
that are equal before pulse arrival. $v = \omega - \omega_0$ is the
detuning between the optical frequency $\omega$ and the trion resonance
frequency $\omega_0$, and $f_{1\pm}(t)$ is proportional to the
smooth envelopes of the circular components $\sigma^+$ and
$\sigma^-$ of the excitation pulse given by
\[
f_{1\pm}(t) = -\frac{2\mathrm e^{\mathrm i \omega t}}{\hbar}\int \mathsf d(\bm r)
E_{\sigma_{1\pm}}(\bm r,t)\mathrm d^3 r\:.
\]
$\Omega_{1\pm}=\sqrt{|f_{1\pm}|^2+v^2}$. The index 1 in $f_{1\pm}$,
$E_{\sigma_{1\pm}}$ and $\Omega_{1\pm}$ indicates the first pulse.
The expressions given above in Eq.~(\ref{eq:eq12}) are for the
elements only that are needed to calculate the FWM signal.

Then we have to consider the dynamics in a transverse magnetic
field. The magnetic field is applied perpendicular to the
propagation direction of the incident light and parallel to the
structure growth axis. Let us first consider the situation without
taking into account spin precession of the hole in trion. Right
before the second pulse arrival we have: \bea
\rho_{13}(\tau_{12}) &=& \rho_{13}(t_p)\cos(\omega_L (\tau_{12}-t_p)/2)\mathrm e^{-(\tau_{12}-t_p)/T_2}\mathrm e^{i\omega_0 (\tau_{12}-t_p)}\nonumber\\
\rho_{23}(\tau_{12}) &=& -i\rho_{13}(t_p)\sin(\omega_L (\tau_{12}-t_p)/2)\mathrm e^{-(\tau_{12}-t_p)/T_2}\mathrm e^{i\omega_0 (\tau_{12}-t_p)}\nonumber\\
\rho_{24}(\tau_{12}) &=& \rho_{24}(t_p)\cos(\omega_L (\tau_{12}-t_p)/2)\mathrm e^{-(\tau_{12}-t_p)/T_2}\mathrm e^{i\omega_0 (\tau_{12}-t_p)}\nonumber\\
\rho_{14}(\tau_{12}) &=& -i\rho_{24}(t_p)\sin(\omega_L (\tau_{12}-t_p)/2)\mathrm e^{-(\tau_{12}-t_p)/T_2}\mathrm e^{i\omega_0 (\tau_{12}-t_p)}\nonumber\\
\label{eq:eq13} \eea Here $\tau_{12}$ is the delay between the two
pulses, $\omega_L$ is the electron Larmor precession frequency, and
$T_2$ is the phenomenological decay time of optical coherence.

After the second pulse action only the following terms can
contribute to the third order trion polarization: \bea \rho_{13}(\tau_{12} +t_p)
&\propto& \frac{(f_{2+}^*)^2
\rho_{31}(\tau_{12})\sin^2(\Omega_{2+}t_p/2)}{\Omega_{2+}^2}e^{i\omega
t_p}
\nonumber\\
\rho_{24}(\tau_{12} +t_p) &\propto&
\frac{(f_{2-}^*)^2 \rho_{42}(\tau_{12})\sin^2(\Omega_{2-}t_p/2)}{\Omega_{2-}^2}e^{i\omega t_p}
\nonumber\\
\rho_{23}(\tau_{12} +t_p) &\propto&
\frac{f_{2+}^*f_{2-}^*\rho_{41}(\tau_{12})\sin(\Omega_{2+}t_p/2)\sin(\Omega_{2-}t_p/2)}{\Omega_{2+}\Omega_{2-}}\mathrm e^{i\omega t_p}
\nonumber\\
\rho_{14}(\tau_{12} +t_p) &\propto&
\frac{f_{2+}^*f_{2-}^*\rho_{32}(\tau_{12})\sin(\Omega_{2+}t_p/2)\sin(\Omega_{2-}t_p/2)}{\Omega_{2+}\Omega_{2-}}\mathrm
e^{i\omega t_p} \label{eq:eq14} \eea The circular polarized components of the
FWM signal are obtained through: \bea
P^+& \propto & \mathsf d^*\rho_{13}\mathrm e^{i\omega_0 t}+c.c.\nonumber\\
P^-& \propto & \mathsf d^*\rho_{24}\mathrm e^{i\omega_0 t}+c.c.
\label{eq:eq15} \eea One can simplify these expressions if the
excitation is weak [$\cos (\Omega_{\pm} t_p) \approx
1-(\Omega_{\pm} t_p)^2/2$ and $\sin (\Omega_{\pm} t_p) \approx
\Omega_{\pm} t_p$], the detuning $v$ is small ($vt_p < 1$) and $t_p
\ll \tau_{12}$.

Then, after excitation by the second pulse we obtain: \bea
P^+&\propto & \frac{it_p^3\mathrm e^{i\omega t}\mathrm e^{iv
(t-2\tau_{12})}\mathrm e^{-t/T_2}}{8}
[(f_{2+}^*)^2f_{1+}\rho_{11}(0)\cos(\omega_L \tau_{12}/2)\cos(\omega_L (t-\tau_{12})/2)\nonumber\\
&+&(f_{2+}^*f_{2-}^*f_{1-})\rho_{22}(0)\sin(\omega_L \tau_{12}/2)\sin(\omega_L (t-\tau_{12})/2)]
\label{eq:eq15}
\eea
\bea
P^-&\propto & \frac{it_p^3\mathrm e^{i\omega t}\mathrm e^{iv (t-2\tau_{12})}\mathrm e^{-t/T_2}}{8}
[(f_{2-}^*)^2f_{1-}\rho_{22}(0)\cos(\omega_L \tau_{12}/2)\cos(\omega_L (t-\tau)/2)\nonumber\\
&+&(f_{2+}^*f_{2-}^*f_{1+})\rho_{11}(0)\sin(\omega_L \tau_{12}/2)\sin(\omega_L (t-\tau_{12})/2)]
\label{eq:eq16}
\eea

To calculate the FWM signal from the ensemble of trions one has to
sum the Eq.~(\ref{eq:eq16}) over all $\omega_0$. If the distribution
function is Gaussian with a central frequency is $\omega$ and a
dispersion $\Delta \omega_0$, this gives the echo signal: \be
P^{\pm} \propto \mathrm e^{-i\omega t} \mathrm
e^{-(t-2\tau_{12})^2\Delta \omega^2_0/2} \mathrm
e^{-t/T_2}[...]+c.c. \label{eq:eq17} \ee Here $[...]$ indicates
other factors.

Finally, the circular polarized component of the amplitude of the echo signal
at $t=2\tau_{12}$ is: \bea P_{pe}^+ &\propto & \mathrm
e^{-2\tau_{12}/T_2}
[(f_{2+}^*)^2f_{1+}\rho_{11}(0)\cos ^2(\omega_L \tau_{12}/2)\nonumber\\
&+&(f_{2+}^*f_{2-}^*f_{1-})\rho_{22}(0)\sin ^2(\omega_L \tau_{12}/2)]
\label{eq:eq18}
\eea
\bea
P_{pe}^-
&\propto & \mathrm e^{-2\tau_{12}/T_2}
[(f_{2-}^*)^2f_{1-}\rho_{22}(0)\cos ^2(\omega_L \tau_{12}/2)\nonumber\\
&+&(f_{2+}^*f_{2-}^*f_{1+})\rho_{11}(0)\sin ^2(\omega_L
\tau_{12}/2)] \label{eq:eq19} \eea The last equations contain the non-zero contributions to the photon echo signal listed in the Table I.
Using Eqs.~\ref{eq:eq18} and \ref{eq:eq19} we obtain the expressions for different polarization configurations of the
two-pulse scheme.

\subsection{Optical polarization for $\sigma^+ \sigma^+$ excitation:}

\bea
P_{pe}^+ &\propto & (f_{2+})^2(f_{1+})^*\mathrm e^{-2\tau_{12}/T_2}[\rho_{11}(0)(1+\cos(\omega_L \tau_{12}))/2]+c.c.\nonumber\\
P_{pe}^-&=&0
\label{eq:eq20}
\eea

\subsection{Excitation by linearly polarized pulses:}

The amplitude of the echo-signal for parallel polarized pulses is
given by: \bea P_{pe}^{lin, co}&\propto & \frac{1}{2}(f_{1})^*(f_{2})^2\mathrm
e^{-2\tau_{12}/T_2} \label{eq:12} \eea Here
$(f_{1+})^*=(f_{1-})^*=(f_{1})^*$,
$(f_{2+})^*=(f_{2-})^*=(f_{2})^*$.

The amplitude of the echo-signal for  orthogonal polarized pulses
is: \bea P_{pe}^{lin, cross}&\propto & \frac{1}{2}(f_{1})^*(f_{2})^2\mathrm
e^{-2\tau_{12}/T_2}\cos(\omega_L \tau_{12}) \label{eq:12} \eea

\subsection{Hole spin precession}

One can also calculate the amplitude of the echo signal at
$t=2\tau_{12}$ taking into account the hole spin precession with
frequency $\omega_L^T$.  For the $\sigma^+ \sigma^+$ excitation
configuration the $\sigma^+$ component of the optical polarization
is: \bea P_{pe}^+ &\propto & (f_{2+})^2(f_{1+})^*\mathrm
e^{-2\tau_{12}/T_2} [\rho_{11}(0)((1+\cos(\omega_L
\tau_{12}))/2)(1+\cos(\omega_L^T \tau_{12}))/2]+c.c. \eea One sees
that the hole spin precession appears here as additional modulation
of the signal. If the decay time $T_2$ is short, so that only a few
oscillations of the electron Larmor precession are seen, this
additional modulation shows up as faster decay.

For linearly cross-polarized pulses the amplitude of the echo signal
at $t=2\tau_{12}$ is then given by: \bea P_{pe}^{lin, cross}&\propto &
(f_{1})^*(f_{2})^2\mathrm
e^{-2\tau_{12}/T_2}\cos((\omega_L-\omega_L^T) \tau_{12})
\label{eq:12} \eea Obviously the precession frequency for linear
excitation differs from the one for circular excitation.

\section{Experimental technique}

The investigated sample comprise 5 decoupled (isolated) CdTe 20~nm QW width separated by 110~nm Cd$_{0.78}$Mg$_{0.22}$Te barriers grown by molecular beam epitaxy. The barriers are doped by iodine donors providing low density electron gas in the QW with $n_e\approx 10^{10}$~cm$^{-2}$ ~\cite{Zhukov06,Bartsch11}.  The sample was mounted into a liquid He bath cryostat and kept under a temperature of 2~K. Magnetic fields up to 0.7~T were applied using electromagnet in Voigt geometry.

We used tunable self mode-locked Ti:Sa laser as a source of optical pulses with the duration of about 2.5~ps and the repetition rate of 75.5~MHz. Degenerate transient FWM experiment was performed in the reflection geometry. Two pulses with practically the same photon energy $\hbar\omega_1=\hbar\omega_2$, non-collinear wavevectors $\mathbf{k}_1$ and $\mathbf{k}_2$, and variable delay time $\tau_{12}$ are focused at the sample into a spot of about $200~\mu$m diameter. The intensities of each pulse were selected to remain in linear excitation regime for each of the beams (pulse energy around 10-100~nJ/cm$^{2}$). The signal was collected in $2\mathbf{k_2}-\mathbf{k_1}$ direction. We used interferometric heterodyne detection where the FWM signal and the reference beam are overlapped at the balanced detector \cite{heterodyne}. For this optical frequencies of pulse 1 and reference pulse were shifted by 40~MHz and 41~MHz with acousto-optical modulators. The resulting interference signal at the photodiode was filtered with high frequency Lock-in amplifier at $|2\omega_2-\omega_1-\omega_{\mathrm{ref}}|=1$~MHz. This provided high sensitivity measurement of the absolute value of the FWM electric field amplitude in real time when scanning the reference pulse delay time $t$, which is taken with respect to pulse 1 time arrival. The polarization of the first and the second pulses as well as the detection polarization were controlled with the help of retardation plates in conjunction with polarizers.

\end{document}